\documentclass[12pt,preprint]{aastex}
\usepackage{epsfig}

\newcommand{\ms}{M$_{\odot}$}
\newcommand{\ls}{L$_{\odot}$}
\newcommand{\mm}{$\,\rm\mu m$}

\newcommand{\hh}{H$_2$}
\newcommand{\kms}{$\,\rm km\,s^{-1}$}
\newcommand{\kkms}{$\,\mathrm{K\,km\,s^{-1}}$}
\newcommand{\kmsm}{\rm{km\,s^{-1}}}
\newcommand{\kmspc}{$\rm km\,s^{-1}\,pc^{-1}$}
\newcommand{\water}{$\rm{H_2O}$}
\newcommand{\be}{\begin{equation}}
\newcommand{\ee}{\end{equation}}
\newcommand{\dvdz}{$\rm dv/dz$}
\newcommand{\s}{$\sim$}
\newcommand{\hnkt}{\frac{h\nu}{kT}}

\newcommand{\as}{$^{\prime\prime}$}
\newcommand{\ccm}{\,\mathrm{cm^{-3}}}  

\newcommand{\too}{\!\rightarrow\!} 
\newcommand{\jone}{\small{$\rm J\eqq1\too0$}\normalsize}
\newcommand{\jtwo}{\small{$\rm J\eqq2\too1$}\normalsize}

\newcommand{\jseven}{\small{$\rm J\eqq7\too6$}\normalsize}
\newcommand{\jfourteen}{\small{$\rm J\eqq14\too13$}\normalsize}
\newcommand{\jsixteen}{\small{$\rm J\eqq16\too15$}\normalsize}
\newcommand{\jeighteen}{\small{$\rm J\eqq18\too17$}\normalsize}

\newcommand{\am}{$^{\prime}$ }

\newcommand{\by}{$\times$}
\newcommand{\e}{\times10^}

\newcommand{\eqq}{\!=\!} 

\newcommand{\sas}{Sgr~A$^*$}

\newcommand{\jj}{\jseven}

\shorttitle{CO \jj\ in the Central 2 pc}
\shortauthors{C.M. Bradford et al.}

\begin{document}

\title{Warm Molecular Gas Traced with CO $J=7\rightarrow6$ in the Galaxy's Central 2~pc: 
Dynamical Heating of the Circumnuclear Disk}

\author{C. M. Bradford\altaffilmark{1}, G. J. Stacey, T. Nikola}
\affil{Department of Astronomy, Cornell University, Ithaca, NY 14853}
\author{A. D. Bolatto}
\affil{Department of Astronomy and Radio Astronomy Laboratory, Universtity of California, Berkeley, CA 94720-3411}
\author{J.M. Jackson}
\affil{Institute for Astrophysical Research, Boston University, Boston, MA 02215}
\and \author{M. L. Savage, J. A. Davidson}
\affil{USRA SOFIA, NASA Ames Research Center, Moffet Field, CA 94035}


\altaffiltext{1}{present address: Jet Propulsion Laboratory, Mail Stop 169-507, Pasadena,
CA, 91109, Matt.Bradford@jpl.nasa.gov}


\begin{abstract}
We present an 11\as\ resolution map of the central two parsecs of the
 Galaxy in the CO \jseven\ rotational transition.  The CO emission
 shows rotation about \sas, but also evidence for non-circular
 turbulent motion and a clumpy morphology.  We combine our dataset
 with available CO measurements to model the physical conditions in
 the disk.  We find that the molecular gas in the region is both warm
 and dense, with T$\sim$200--300~K, $\rm n_{H_{2}}\sim
 $5--7$\times10^{4}\,cm^{-3}$.  The mass of warm molecular gas we measure in
 the central two parsecs is at least 2000~\ms, about 20 times the
 UV-excited atomic gas mass, ruling out an UV heating scenario for the
 molecular material.  We compare the available spectral tracers with
 theoretical models and conclude that molecular gas is heated with
 magneto-hydrodynamic shocks with v$\sim$10--20~\kms and
 B$\sim$0.3--0.5~mG.  Using the conditions derived with the CO
 analysis, we include the other important coolants -- neutral oxygen
 and molecular hydrogen -- to estimate the total cooling budget of the
 molecular material.  We derive a mass to luminosity ratio of $\sim$2--3~\ms~/~\ls, which is consistent with the
total power dissipated via turbulent decay in 0.1~pc cells with $\rm v_{rms}\sim
 15\, km\,s^{-1}$.  These size and velocity scales are comparable to the observed clumping scale and the
 velocity dispersion.  At this rate, the material near \sas\ its dissipating
 its orbital energy on an orbital timescale, and cannot last for more
 than a few orbits.  Our conclusions support a scenario in which the features near \sas\ such as the CND and northern arm
are generated by infalling clouds with low specific angular momentum.
 \end{abstract}


\keywords{Galaxy: center --- Galaxy: nucleus -- ISM: molecules -- turbulence}

\section{Introduction}\label{sec:intro}
The Galactic Center provides the opportunity to study many of the
fascinating phenomena associated with massive black holes and galactic
nuclei in the greatest detail.  The region is likely similar to
the nuclei of galaxies beyond the Milky Way, particularly those harboring low-luminosity AGN (LLAGN) with similar
sub-Eddington luminosities. 
Since the Galactic Center radio and infrared sources were first discovered
decades ago \citep{or60,bn68,bn69,bb74}, the central source and its
circumnuclear material has been the subject of intense study from
infrared to radio freqencies using both continuum and spectrosocpic
probes.

In the radio continuum, the central few parsecs show a spiral-like
structure of thermal emission, termed the minispiral or Sgr A West,
roughly centered on the bright, non-thermal point source \sas\
\citep{eke83,lc83}.  The arms of the radio minispiral are interpreted
as the ionized remains of tidally-disrupted in-falling clouds, a
notion supported by the large ($\sim$ 500 \kms) velocity dispersion in
the ionized gas observed via mid-IR fine structure and radio
recombination lines \citep{lac80,sbv89}.  The distribution of
molecular and neutral atomic gas shows a different morphology --
interferometric observations of HCN show a substantial portion of an
inclined ring-like structure of diameter approximately 3~pc
surrounding \sas, with kinematic evidence for rotational motion of
about 110~\kms around the center \citep{gus87,wmb89}.  In contrast, the neutral oxygen
emission shows a peak inside the circumnuclear ring, but also shows
general evidence of rotational motion in the same sense as the HCN. \citep[][henceforth J93]{jac93}.

The first far-IR observations demonstrated that the substantial IR
luminosity (few $\rm \times 10^{6}\,\rm L_{\odot}$) at the center of
the galaxy is due to dust absorbing UV and optical energy and
reradiating it at longer wavelengths \citep{hfe71,gat77,gb81}.
Subsequent mapping revealed bi-lobal structure with temperature
decreasing with distance from the center, suggesting a circumnuclear
ring or disk around a central 2--3 pc evacuated cavity containing the
primary luminosity source \citep{bgw82}.  

The far-IR, radio continuum, and molecular features have been unified with high-resolution far-IR images \citep{lat99}
Far infrared images at 31 and 38 \mm\ provide color temperature, optical depth, and luminosity maps of the inner
$2\times3$ pc region, and trace the deposition of UV
luminosity and constrain the masses of the features.  The most prominent feature is the CND itself.  Here the far-IR typically peaks 1--3\as\
(0.03--0.1 pc) further from \sas than the radio continuum (e.g. Sgr A West), and the molecular gas ring as traced with
HCN has a larger radius still.  This is as would be expected in a typical photodissociation region (PDR) situation in
which the progression from the UV source is: ionized gas traced with radio continuum---photodissociated gas and warm
dust---molecular gas.   Given the similarity in the morphology and kinematics of the the radio arcs, the far-IR ring, and
the HCN ring, a plausible scenario is therefore that these features trace different aspects of the same
structure, a ring or torus of material orbiting the central mass, called the circumnuclear disk (CND).    

In addition to the CND, the other other prominent features distinctly traced in both the far-IR and radio are the
Northern Arm (NA) and the east-west bar; both part of the radio source identified as SGR A West.  The radio
continuum and far-IR morphologies again make a nice match, and their kinematics as probed via the radio
recombination lines suggests that these features are distinct infalling streamers
on parabolic orbits around \sas.  The neutral oxygen emission ([OI]) was originally attributed to material
within the cavity formed by the CND (J93), but has been shown to be primarily originating from the NA itself.  
The sizes and column densities as traced with the dust continuum imply average gas densities of 1.6--4$\e4\ccm$,
consistent with the average densities derived for the atomic gas (10$^4$--10$^5\ccm$).   

Aside from a few local heating sources, the color temperature map of the region presented by L93 is strongly peaked
in the center near \sas, and the run of temperature with radius follows the same $T\sim r^{-0.4}$
law across all of the features.   The central cluster massive young He I stars centered within an arcsecond of \sas\
(collectively called IRS 16), in concert with the other stellar clusters thus makes an appealing mechanism for
centrally ionizing and heating the material in the CND and the other streamers (L99).

While the UV sources near \sas\ are thus a good explanation for the warm
dust and atomic gas in the NA, and the inner edge of the CND, UV photons are not necessarily
the heating source of the molecular material that comprises the bulk
of the CND.  Various theoretical models for the formation of the CND
invoke infalling clouds and require cloud-to-cloud collsions
\citep{vd02,san98}.  Dynamical heating has been suggested as a
mechanism for molecular clouds in the Galaxy's central 500~pc to
explain bright rotational \hh\ lines \citep{rf00,rf01} and the
presence of highly-excited NH$_{3}$ \citep{wil82,hh02}.  At the position of \sas\
itself, the ro-vibrational molecular hydrogen lines show an almost
completely thermal spectrum, much like the Orion-KL shock
\citep{tan89,gat86}, demonstrating that UV is not responsible for the
heating of this warm molecular gas.

Thus, the question remains open: what are the relative roles of the
dynamical and ultraviolet energy sources in exciting the molecular
gas in the central 2~cpc?  More generally, what is the origin, lifetime, and eventual fate
of this material?  To address these
questions, we have mapped the \jj\ rotational transition of carbon
monoxide (CO) in this region.  The mid-J CO rotational transitions are
especially useful probes for the Galactic Center because they are
thermalized and optically thick in the circumnuclear disk, but suffer
little extinction and virtually no line optical depth in the
foreground Galactic material.  We use CO lines to estimate the mass and conditions of the
warm molecular gas.  We trace more than 2000\ms\, and derive a
moderate density, $\rm n\sim 2-7\e{4}\,cm^{-3}$, and warm temperature
$\rm T\sim 200-300\,K$.  Unlike the atomic gas and the warm dust
traced in the far-IR continuum, the bulk of this warm molecular
component is likely heated dynamically via magneto-hydrodynamic shocks
due to clump-clump collisions, not UV photons from the central
sources.

The observations are described in Section~\ref{sec:observations}.
Section~\ref{sec:results} presents the results of our CO observations
and a brief comparison with the morphological and kinematic features
observed in other tracers.   (We will present a more detailed morphological study using a larger,
higher-spatial-resolution map in a subsequent article.)   Analysis of the CND molecular gas
excitation is presented in Section~\ref{sec:phys-cond-cnd}, and a
discussion of the CND gas heating mechanism follows in
Section~\ref{sec:heating}.   

\section{Observations}
\label{sec:observations}

The CO \jj\ ($\nu=806.65\rm, GHz$) observations were made in April
1999 at the James Clerk Maxwell Telescope (JCMT) on the commissioning
run of the submillimeter Fabry-Perot spectrometer SPIFI.  Details of
the instrument and its performance can be found in \citet{bra02}.  In
its first-light configuration, the detector array included 12
bolometers, arranged on an incomplete 4$\times$4 grid with 8\as\
spacing.  Over the two nights, the array was placed in eleven
positions, six of which included sufficient field rotation to warrant
being split into two pointings, generating a total of 17 independent
pointings of the array.  Each pointing is comprised of 30--50 spectral
scans of the Fabry-Perot with the telescope nodded every two scans to
exchange the source and reference beams.  For these observations, the
velocity resolution of the spectrometer was 70\kms, and the total scan
bandwidth was 620\kms.  The spectra were sampled at 50\kms, but the data are rebinned at a finer resolution to allow
proper co-adding of the various pixels around the array, each of which has a small relative velocity shift. 

Calibration was based on Mars, which was near opposition and was taken
to be a uniform disk of size 15.8\as\ and brightness temperature
201~K, the mean of the models of \citet{wri76} and \citet{rud87}.  At
the time of the observations, the forward coupling of the telescope
was measured to be 0.65, of which 8.5\% coupled to Mars.  This low
value of the coupling to Mars resulted from a combination of telescope
surface and misfigured reimaging lenses.  On subsequent observing runs
this coupling was measured to be $\sim$ 20\%, consistent with other
observations at 800 GHz.

All Galactic Center observations were conducted in submillimeter band
2 weather ($\tau_{225}>0.05$).  When combined with the low elevation
of the source, this provided a transmission to the source between 0.03
and 0.10.  The atmospheric transmission for each scan was determined
with the 225 GHz radiometer according to 

\begin{equation} \rm
\tau_{806\, GHz} = 37 \,\left(\tau_{225 \,GHz} - 0.012\right).
\label{eq:2} 
\end{equation}  

This relation was calibrated by measuring sky emission against an
ambient load as a function of zenith angle, and comparing the derived
opacity to the measured 225 GHz opacity.  With the low telescope
coupling and atmospheric transmission, the system NEFD for the
observations was $\sim \rm 3\times
10^{-21}\,erg\,s^{-1}\,cm^{-2}\,Hz^{-1}\,Hz^{-\frac{1}{2}}$.


Once the twelve spectra were generated from each pointing, the total
flux for each pixel in each pointing is calculated by integrating over
$\rm -160\,km\,s^{-1} < v < 160\,km\,s^{-1}$, to provide adequate baseline, and trace the same gas as with HCN and
NH$_{3}$.  The overlaps in the various pointings then allow for determining the appropriate scalings
between the various pointings, these relative scalings are as much as
a factor of 1.8 over the entire set of 17 pointings, but the scaling
does not change by more than 30\% between two pointings adjacent in
time.  

The spectra were then mosaiced together by regridding onto a uniform
0.43\as\ spatial by 10\kms\ spectral grid.  At a given velocity, $\rm
T_{MB}$ in each pixel is determined by any surrounding data points as
follows: Each observed point is placed into the spatial array as a
delta function with magnitude equal to its intensity.  The map is
generated by smoothing this array with an 11\as\ gaussian kernel, then
normalizing each position by a template map obtained by the same
smoothing of an array with unity at each data point.  With the incompletely-filled array, and the effects of
field rotation some
regions of the map are highly sampled, other are very sparse, thus the
sensitivity in the map varies substantially.  For a typical interior
point in the map, there are between one and three spectra per beam,
and we estimate the relative uncertainty to be $\sim$~300\kkms\ from
beam to beam, and about twice this value across the entire map, the
additional error accumulating due to the mosaicing together of several
pointings.  The map boundaries are defined to be where the template
array drops below 0.4, that is, where there is not a single observed
point within 6\as, thus large regions near the map boundaries may be
influenced only by a small number of observed points.  The overall
calibration uncertainty is estimated to be about 30\%, due to the
varying atmospheric conditions during the observations and
calibration.





\section{Results}
\label{sec:results}

The integrated intensity map in CO \jseven\ of the central 2~pc is presented in Figure~\ref{fig:gc}, along with five
spectra of various positions offset from \sas.  Emission is observed throughout
the 2\am$\times$ 1\am\ mapped region, with a minimum integrated
intensity of 900~\kkms (referred to the main beam temperature).  The
peak in the CO emission is in the southwest, likely part of the CND, with an
intensity of 5400~\kkms, and the average intensity throughout the
region is 2550~\kkms.  In this section we first compare our results
with previous observations of the same tracer.  We then compare our
map with high-resolution maps in the far-IR continuum and the density
tracer HCN in order to address the morphology and kinematics of the
warm-CO containing gas.  Hereafter we adopt a distance of $7.9\pm0.3$~kpc to
the Galactic Center, according to the analysis of \citet{mcn00}.

\subsection{Comparison with previous observations}
\label{sec:compare}

The integrated intensities we measure are somewhat lower than those of
\citet{har85}, who observed CO \jseven\ in nine 30\as\ beams offset in
galactic longitude from \sas.  \citet{har85} do not present a two-dimensional map, but their brightest observed
position is toward the southwest part of the Circumnuclear Disk (CND) with a total measured luminosity of
200~\ls\ in the 30\as\ beam.  Integrating over a 30\as\ beam centered on the peak
in our map results in only 105~\ls.
A potential explanation for this discrepancy is 
self-chopping in our observation.  We employed a 120\as\ chop in
RA, the maximum usable for efficient operation at the JCMT.  With this
chop the reference beam does approach the ``+20~\kms'' and
``+50~\kms'' Sgr A clouds, large extended clouds lying roughly
1--3\am\ to the east of \sas\ \citep{gen90}.  \citet{gen90} have
observed these objects in CO $J\eqq7\too6$ with a roughly eight 30\am\
pointings, finding peak intensities around 600--800~\kkms, over large
($\sim$1\am) regions.  \citet{har85} were able to use a 190\as\ chop
throw on the IRTF, which clears the majority of the Sgr~A clouds.
While the chopping onto these clouds is thus a difference between our
observations and those of \citeauthor{har85},
the effect is not likely to account for the significant difference in
observed intensities.  Because we were nodding the telescope,
exchanging between an eastern and a western reference beam, the error
introduced by an 800~\kkms\ source in one of the reference beams is at
most $\sim$ 400~\kkms, or $\sim$ 10\% of the observed flux density.
Our spectra show no obvious indication of reference beam emission and
a comparison of our primary dataset with a small sample dataset
obtained using a 90\as\ chop does not show any evidence of reference
beam emission affecting the map.  Apart from the overall calibration
difference, our map corresponds well with the large-beam pointings of
\citeauthor*{har85}, both in relative integrated flux, and in velocity
spread of the emission across the region.

\citet{mcp01} present a single \jseven\ spectrum toward \sas, obtained
with the CSO, their data show an integrated intensity of 700~\kkms\ in
T$_{A}^{*}$.  Taking a main beam efficiency of 0.3, as in our
measurements with the JCMT, this corresponds to a main beam intensity
of 2330~\kkms, 64\% of the 3650~\kkms\ we measure.  Our measurements
are therefore intermediate between the other published values, which
vary by a factor of more than two.

\subsection{Morphology and kinematics} 
\label{sec:morphology}

CO \jseven\ emission is detected throughout our observed region.  We reserve the detailed kinematic and morphological 
study for the higher-spatial-resolution, larger area dataset presented in a subsequent paper.  For the purposes of this
study, we simply overlay the CO dataset on the 38\mm\
continuum image \citep{lat99} in Figure~\ref{fig:overlayir} and the
interferometric HCN map \citep{gus87} in Figure~\ref{fig:overlayhcn}.  We highlight two points which lead to the
conclusions that the gas tracing the mid-J CO bears some relation but is not identical to the CND material traced with
HCN or the high-excitation NH$_{3}$. 

1) The bright CO emission peak in the southwest is coincident with the far-IR peak at the same location:  (-15, -30). 
This location corresponds to a smaller radius (1.2 pc projected) but the same position angle
as the HCN peak.  The bright CO emission 20--30\as\ south of \sas\ is
similar to that found in the map of NH$_{3}$~(6,6), tracing molecular gas at a
few hundred K \citep{hh02}.  We note that with our moderate spatial resolution and relatively small coverage, the CO and
NH$_{3}$ maps do not agree in detail, as the NH$_{3}$ peak is further east.  In
general, both tracers show that warm gas is interior to the circumnuclear ring as traced
with HCN.   

2) Kinematically, the warm CO shows evidence of motion
about \sas.  The spectra in Figure~\ref{fig:gc} reveal a clear
north-south velocity gradient with the same sense of rotation as is
observed in the HCN.  However, the CO \jseven\ does not show the
obvious ring shape, nor the well-organized circular orbit of the dense
gas traced with the HCN.  Neglecting the morphological assymetry, the
observed orbital velocity of $\sim$60~\kms\ line of sight (65~\kms\
deprojected with 67$^{\circ}$ inclination), at a distance of 35\as\,
suggests an interior mass of only $1.3\e6\,M_{\odot}$.  This is
substantially less than the mass traced with HCN, and dramatically
less than the central dark mass traced with the stellar velocity field
(M$\sim$4.0$\e6$ \ms \citep{ghe03,sch03}).  Evidently, the
CO-containing gas does not have angular momentum for a stable orbit at
its distance from the central mass.  \citet{hh02} reached a similar
conclusion in their mapping of NH$_{3}$~(6,6).  Again aside from this general similarity, the two tracers do not share
detailed kinematics:  the CO shows a sense of rotation similar to that of the HCN, while the NH$_{3}$ does not.  Given

\section{Physical conditions in the GC molecular gas} 
\label{sec:phys-cond-cnd}

Bright CO \jseven\ emission implies the presence of both warm and
dense gas.  The $J\eqq7$ level lies 155~K above ground and the
\jseven\ transition has a high critical density, $n_{\rm crit} =
4\e5\ccm$ for thermalization.  Our peak brightness temperature of 33~K
originates in gas with a minimum physical temperature of
$T=\frac{h\nu}{k}[\ln(\frac{h\nu}{k T_{\rm MB}} + 1)]^{-1} = 51
\rm\,K$, if the transition were optically thick, thermalized, and
filled the 11\as\ beam.  This is a strong lower limit -- in reality the gas is at lower density and does not fill the
beam, which requires a higher temperature.


\subsection{LVG radiative transfer calculations} 
\label{sec:lvg}

We combine our \jseven\ observations with published CO observations in
the far-IR (\jfourteen\ and \jsixteen\, \citep{lug87}) and millimeter
(\jtwo\ and \jone\ \citep{ser86}) to estimate the conditions in the
molecular gas in the framework of an LVG radiative transfer model.  The code
calculates intensities of the rotational CO lines up to \jeighteen,
given a molecular hydrogen density, temperature, and velocity gradient
in $\rm{km\,s^{-1}\,pc^{-1}}$.  The process was originally employed
for the study of molecular clouds by \citet{ss74} and \citet{gk74},
and further description of the model used here can be found in
\citet{bra03}.  As in their study of the nuclear molecular gas in
NGC~253, we adopt a velocity gradient much larger than the traditional
values of a few \kmspc\ appropriate only for cool, quiescient
molecular clouds in the disk of the Milky Way.  For the material in the central 2~pc, we take \dvdz\ to be
300 $\rm{km\,s^{-1}\,pc^{-1}}$, consistent with our observed velocity width of $\sim$100\kms\ in an
11\as\ beam.  We adopt $8\e{-5}$ for $X_{\rm
CO}$, the CO fractional abundance relative to \hh.  This values is intermediate between that inferred
from observations of the massive Galactic clouds Sgr B2 and Orion IRc2 (for which $X_{\rm CO} \sim 6\e{-5}$
\citep{lhh84,elc91}), and that of
chemical models (for which ${X_{\rm CO}
\sim 1.5\e{-4}}$ (e.g. \citet{bla87})).  Collisional rates are from
\citet{flo01,flo04}, increased by 21\% to account for collisions with
helium \citep{mck82,vc88,fl85,sch85}.  

For a meaningful run of line intensity with J, all observed
intensities are averaged over a 44\as\ beam corresponding to the
far-IR beam size.  This average beam is centered toward the \jseven\
peak in the southwest at (-15,-30).  Figure~\ref{fig:gclvg} shows the
constraints imposed by two line ratios.  We use the \jsixteen\ to
\jseven\ and \jseven\ to \jtwo\ ratios, as these straddle the peak in
the run of intensity verses J.  The heavy contours correspond to
observed ratios, and their intersection region represents the range of
possible solutions.  With errors on the submillimeter and far-IR
intensities estimated at 30\%, the line ratios should be considered
good to $\sim$ 40\%, admitting a large range of solutions over which
temperature can be traded with density.  It is clear that the
temperature is greater than 200 K, and that the molecular hydrogen
density is less than $10^{5}\,\rm cm^{-3}$, though there are no
stringent upper limits on the temperature using the available rate
coefficients.  We favor temperatures toward the lower end of the allowed range shown in Figure~\ref{fig:gclvg} 
because of the lack of evidence for higher-temperature molecular tracers, as discussed in 
Section~\ref{sec:luminosity}).  Given this constraint, we plot a suitable solution ($\rm
n_{H_2} = 7.1\e4\ccm$, $T=240\, K$) superposed on the observed line
intensities in Figure~\ref{fig:cofit}.
In this model, $\tau$ exceeds unity for ${J_{\rm upper}} =$~7--10, but
does not exceed 2 due to the large velocity gradient appropriate in
the CND.  Our conclusions for the SW CND are inconsistent with the conditions
reported by \citet{mcp01} for the gas toward Sgr~A$^*$ itself: $\rm T\sim
250 K$, and $\rm n_{H_2}\sim 3\e5\,cm^{-3}$.   We deduce smaller densities, in spite of using the same collisional rates from
\citet{sch85}.  All of our inferences are based on a
fit to the observations toward the peak of the CO emission in the southwest CND.  It is possible that the
molecular gas is more highly excited toward \sas\ itself than in the SW CND,
though distinguishing between the two is difficult given the large beams for the high-J far-IR lines.   Based on the
50\as\ resolution ISO maps of \sas\ in CO
\jfourteen\ \citep{whi98}, we estimate that the CO line ratios are
similar throughout the region, suggesting similar excitation conditions in the gas.  
If so, the discrepancy may arise from our large velocity gradient which
should be applicable the given the dynamics of the region, which allows the J=7 transition to
radiate more efficiently and with lower collisional excitation than in
more optically thick models.  
\subsection{Molecular gas mass, clumping}
\label{sec:mass}

Using the model plotted in Figure~\ref{fig:cofit}, the required CO
column density is $\rm N_{CO} = 1.2\e{18}\, cm^{-2}$ in the 44\as\
beam, or $\rm N_{H_{2}}=1.5\e{22}\,cm^{-2}$ with our assumed CO
abundance of $8\e{-5}$.  The 44\as\ represents 2.2 pc$^{2}$, and total
molecular gas mass in this beam is therefore 530~\ms.  If the excitation conditions are similar throughout our mapped
region, then we can scale the column density with the observed \jseven\ intensity.  In this case, the beam-averaged
column density in a beam toward \sas\ itself is $\sim$60\% of our peak value, and is 6 times that 
reported by \citet{mcp01}.    This discrepancy is difficult to reconcile.   It could be due in part to
the larger CSO beam relative to our JCMT observations, though this effect is unlikely given that our map
shows extended emission.  
More fundamentally, even if the \jseven\ transition were optically thin and thermalized, it would be impossible
to generate the observed emission toward the peak in our map with the
CO column density reported by \citet{mcp01}.  On the other hand, our
analysis could be consistent with a larger CO column given the lack of any optically thin mid-J CO isotopic transitions. 
Other analyses
of cooler material suggest somewhat larger molecular column densitites.  For example J93 compile low-J CO, CS, and HCN
measurements, and infer $\rm N_{H_{2}} = 4\e{22}-4\e{23}\,cm^{-2}$, in 10\as--30\as\ beams
toward the SW CND.  They suggest that the HCN emission comes from very
dense ($\rm n_{H_{2}}=10^{6}-10^{8}\,cm^{-3}$) material, distributed
with low beam filling factor in small clumps.  While our analysis
indicates that this density is not appropriate for the material which
produces the bulk of the mid-J CO emission, it is possible that clumps at
these high densities could harbor molecular gas which could be locally
optically thick, and thus inefficient in contributing to the total CO
emission.   Again, we stress that
our model employs a higher temperature, and larger velocity gradient,
both of which allow the CO to radiate more efficiently than previously
considered.   Thus our mass and column density estimates should be considered as lower limits to the total warm
molecular gas mass.  Finally, we note that our analysis is not sensitive to cool gas which
does not radiate in the high-J lines, and a cool component could exist along with the warm one we measure.  Such a
two-component molecular medium (25\% at 200~K, 75\% at 25~K) has been proposed based on the high-J NH$_{3}$ observations
of GC clouds \citep{hue93,hh04}.

Scaling the mass in the 44\as\ beam to the entire map
using the CO~\jseven, intensity yields a mass of $\sim 2\e{3}$~\ms\ as a lower limit to the warm molecular gas in the
central 2~pc.
 Comparing the total mass with the derived density implies a volume
filling factor of $4.4\e{-2}$, and an area filling factor of 0.12 in
the 44\as\ beam.  This filling factor is reasonable -- it is
approximately equal to the product of the filling factor of the
overall ring shape as observed at the 11\as\ resolution in the 44\as\
beam used for the calculations, and the intrinsic clumpiness in the
ring on scales below our 11\as\ beam.
The size of the clumps is not constrained with the large-beam
analysis.  At a maximum, all of the 530~\ms\ of gas is in a single
0.6~pc diameter clump, corresponding to $\rm A_{V}=138\, mag$.  The
facts that we are resolving the ring shape in our 11\as\ beam, and that the
HCN and far-IR maps show clumping at the limit of their resolution
(4--12\as~$=$~0.15--0.46 pc) suggest that material is clumped
on smaller scales (L99, \citet{gus87,wri01}).  A clump size of order 0.2~pc might be expected on
physical grounds, as it is comparable to the Jeans length scale, over
which self-gravity becomes important,
\begin{equation} L_J =
\left(\frac{k\,T}{m\,G\,\rho_m}\right)^{\frac{1}{2}} \sim 0.2\, \rm pc
\end{equation}
for the cooler conditions inferred from the CO lines, $T=200\,\rm K$,
and $\rm n_{H_{2}}=7\e4\ccm$.  With this size, the 44\as\ beam harbors
25 clumps, each with a mass of 22~\ms, and the clump column density
corresponds to $\rm A_{V}=47$.  This is sensible in a scenario in
which clumpy molecular material is falling toward \sas\ from greater
distance, but is not appropriate for a steady-state motion at the
distance of the CND.  Any clump with density less than $\rm
n_{H_{2}}\sim6\e{7}\,cm^{-3}$ is Roche unstable in the presence of the
tidal forces of the central mass concentration.  Thus any clumps
associated with the material emitting CO are most likely transient
structures.

\subsection{Molecular gas luminosity}
\label{sec:luminosity}

The CO excitation model provides a good estimate of the mass and total
cooling due to CO in the molecular gas. The above fit to the data
corresponds to a total emergent intensity in the CO lines of
$1.5\e{-2}\,\rm erg\,s^{-1}\,cm^{-2}\,sr^{-1}$ averaged over the
44\as\ modeled region.  This intensity is substantial, comparable to
that of the far-IR atomic fine structure lines (see
Table~\ref{tab:lines}), and about 0.1\% of the integrated dust
continuum intensity.  The modeled column density corresponds to a mass
density of $\rm 7.05\e{-2}\,g\,cm^{-2}$, and assuming that the CO
lines radiate isotropically, we derive a mass to luminosity ratio
of $\rm L_{CO}/{M_{H_2}} = 1.4\,L_{\odot}/M_{\odot}$ for the CO cooling
around \sas.

For a complete energy budget of the warm molecular gas, we now
consider the other coolants.
Detailed chemical studies indicate that in addition to CO, the most important cooling
paths for densities above $10^3\,\ccm$ and
temperatures above 150 K are the rotational transitions of \hh\ and
\water, and the fine structure lines of atomic oxygen ([\ion{O}{1}])
\citep[and references therein]{gl78,nlm95}.  In what follows, we estimate the efficiency of these coolants conclude that
for the conditions we derive with our LVG analysis, it is likely that
CO, \hh, and [\ion{O}{1}] contribute comparably to the cooling budget.

The quadrupole \hh\ lines have low radiative rates and are thus
optically thin and thermalized at the our modeled densities.  The cooling per
gram is then easily estimated using (for details see \citet{lb99}):
\begin{equation}
\rm \frac{L}{M} = \sum_{J} \frac{f_u\, A_{u,l}\, h\nu}{1.43
\cdot 2m_p}
\end{equation} 
where $\rm f_u$ is the fraction in the upper level, given by the
Boltzmann's equation and $\eta_{op}$ is the ortho--to--para ratio.
The rotational \hh\ lines are thermalized for densities above $\rm
10^{4}\, cm^{-3}$, and combined ortho and para cooling
function is not strongly dependent on the $\rm \eta_{op}$ for $\rm
T>100\,K$.  With $\rm T=240\,K$, and $\eta_{op}=3$, the most important
line is the $\rm J=3\rightarrow1$, 
which
produces $\rm 0.29\,L_\odot / M_{\odot}$.  All of the \hh\ lines
combined generate $\rm\sim 0.41~L_\odot/M_\odot$ at this temperature,
some 30\% of the CO cooling.  We do note that the \hh\ lines become
much more efficient coolants as the temperature increases.  For $\rm
T=325\,K$, the \hh\ line cooling becomes comparable to our measured CO
cooling, and any moderate density ($\rm n<10^{6}\, cm^{-3}$), high
temperature regions will be cooled primarily by the \hh\ lines.


As the density increases, the oxygen-bearing species O, O$_{2}$, and
\water\ become important.  Because the oxygen elemental abundance
($\sim3\e{-4}$) is larger than that of carbon ($\sim1.4\e{-4}$)
\citep{ss96}, there is atomic and molecular oxygen in the UV-shielded
cloud interiors where all the gas-phase carbon is locked up in CO.
The chemical models for warm molecular gas (e.g.\citet{nlm95}) predict
a O$^0$ and O$_{2}$ abundances relative to \hh\ of $1\e{-4}$ (each)
for $\rm T<250\,K$.  As the temperature increases to $\sim$ 500~K, all
of the oxygen not locked in CO is likely in the form of \water; its
contribution is described below.  Since O$_{2}$ is not an important
coolant due to its lack of a strong dipole, we only need to consider
the fine structure transitions of [OI], and of the two, only the
\jone\ 63\mm\ is energetically important.  The upper level of the
63\mm\ transition lies 228~K above ground, and would thus would be
well-populated given the temperature in the molecular gas if not for
the high density required to thermalize the transition, $\rm n_{crit}
= 8\times 10^{5}\left(\frac{100}{T [K]}\right)^{0.69}\,cm^{-3} =
4.4\e{5}\,cm^{-3}$.  Neglecting the upper (J=0) level, we can model an
emergent intensity under a two-level approximation:

\begin{equation}
I_{\mathrm [OI]\, 63\rm \mu m}  \sim \frac{h\nu\,N_{\rm O}\, A}{4\pi}\times
\left[\frac{\frac{g_u}{g_l}e^{-\hnkt}}{1+\frac{g_u}{g_l}e^{-\hnkt}+\frac{n_{\rm crit}}{n}}\right] =
1.0\e{-2}\,\rm erg\,s^{-1}\,cm^{-2}\,sr^{-1}  \label{eq:oi}
\end{equation} 

where $g_u$ and $g_l$ are the statistical weights, ${A=9\e{-5}\,\rm
s^{-1}}$, and we have taken the temperature and density from the CO
analysis.  We have assumed that the emission is optically thin, which
is reasonable given the large velocity dispersion of the warm
molecular gas:

\begin{equation}
\tau=\frac{c^3 A N_O}{8\pi\nu^3\Delta v}\left[\left(1+\frac{n_{\rm crit}}{n}\right)e^\hnkt
-1\right] \times
\left[\frac{\frac{g_u}{g_l}e^{-\hnkt}}{1+\frac{g_u}{g_l}e^{-\hnkt}+\frac{n_{\rm
crit}}{n}}\right] = .25 .
\end{equation}

Correcting for extinction and averaging the [OI] maps of J93 over the
44\as\ beam toward the SW peak , \citet{whi98} obtain an intensity of
$2.4\,\rm erg\,s^{-1}\,cm^{-2}\,ster^{-1}$ (see
Section~\ref{sec:atomic}).  In this region, therefore, the neutral
oxygen present in the molecular gas produces nearly half of the
observed [OI] 63\mm\ emission, and represents 70\% of the total
cooling in the CO lines.  Evidently the neutral oxygen plays an
important role in cooling this warm dense molecular gas, and much of
the observed [OI] should be attributed to the molecular phase, not the
atomic gas.

For very warm molecular gas $\rm T>300\,K$, \water\ becomes an
important molecular coolant.  Chemical models predict a \water\
abundance relative to \hh\ as high as $3 \times 10^{-4}$ as the
temperatures increases to 500~K, as it is the dominant oxygen
repository under these conditions.  Once formed through the chemical
network, there are several far-IR rotational transitions of \water,
and their radiative rates are fast because of the molecule's large
dipole moment.  For densities sufficient to thermalize these
transitions, they will therefore dominate the cooling of the gas.  The
critical densities, however, are of order a few $\times 10^5\,\ccm$,
larger than the density inferred with the CO lines.  For the gas we observe,
both the temperature and the density of the gas are thus insufficient
to generate and excite \water\ substantially and it is not likely a
major contributor to the cooling budget.  Recent observations with the
Odin satellite confirm that the cooling is small: $\sim60$~\kkms in
the 557 GHz ($1_{1,0}\rightarrow1_{0,1}$) in the 2.1 arcmin diameter
beam \citep{san03}.  Taking the warm gas to fill the beam gives a
lower limit to the intensity of $\rm
1.1\e{-5}\,erg\,s^{-1}\,cm^{-2}\,sr^{-1}$, about 80 times less than
the CO \jseven\ line.  If all this flux is referred to the 44\as\ beam
used in the CO analysis, the integrated intensity would be
$8.3\e{-5}\,\ccm$, this represents a generous upper limit, but is still
10 times smaller than the observed CO \jseven\ line.
Furthermore, shorter wavelength \water\ lines observed with the 
ISO-LWS are only found in absorption from foreground cold gas in the
continuum -- none of the profiles show evidence for emission with
intensities comparable to the CO lines \citep{mcp01}, and are
therefore rule out a large amount of gas with temperatures higher than
T$\sim$ 250~K.

The theoretical modeling and the observational evidence therefore
support the following cooling budget for the warm molecular gas if it is at the
modeled temperature and density: the CO lines account for half of the
total cooling, the [OI] 63~\mm\ line one third, and the \hh\
rotational lines one sixth.  The total corresponds to a
mass-to-luminosity ratio of 2.7~$\rm L_{\odot}/M_{\odot}$.  As discussed above,
for any material at higher temperatures the \hh\ and \water\ lines will become
more efficient at cooling the gas, and the total cooling of the
molecular gas will substantially increase.



 \subsection{Relationship with the atomic gas}
\label{sec:atomic}

In order to understand the relationship between the atomic and
molecular ISM phases, it is necessary to distinguish between the
morphological features in the central 2~pc.  Estimates of the mass of
atomic gas by J93 were based on their model fits to the [OI] and [CII]
intensities toward the peak 25\as\ north east of \sas.  Most of this
emission is now believed to come from the northern arm (NA), a
tidally-stretched infalling streamer that is dynamically distinct from
the rest of the circumnuclear material (L99, citet{ryg96}).  The dust optical depth
in the NA corresponds to a hydrogen column density of $N_{\rm H} \sim
0.7$--$1.3\e{22}\ccm$ (in the 4\as--6\as\ far-IR beams), corresponding
to an ultraviolet attenuation of only 2--4 magnitudes through the
streamer, insufficient to shield the bulk of the material from
photodissociation.  This optical depth, together with the proximity to
the UV sources makes the NA the dominant feature in the atomic
tracers, but it is negligible in the overall mass budget of the molecular material in the central 2~pc.


By comparison, the dust emission toward the molecular peaks in the SW
CND is associated with cooler, but much more massive structures.  The
optical depths in the clumps in the ring are an order of magnitude
greater than in the northern arm, in general sufficient to shield the
UV from the bulk of the material.  As discussed in
Section~\ref{sec:luminosity}, the warm molecular material as probed
with the CO emission is responsible for much of the [OI] emission
observed toward the SW ring.  The other half arises in the atomic
component.  To estimate the atomic gas mass fraction, we use the [OI]
map to scale the mass estimate of J93 for the NA to the atomic
fraction in the SW CND peak.  Including the extinction correction for
J93, they estimate $\rm N_{H}=1.2\times10^{22}\,cm^{-2}$ based on an
intensity of $\rm 5.0\e{-2}\,erg\,s^{-1}\,cm^{-2}\,sr^{-1}$ (22\as\
beam).  The $\sim 2.0\e{-2}\rm\, erg\,s^{-1}\,cm^{-2}\,sr^{-1}$
observed in the 44\as\, multiplied by 0.5 to remove the molecular gas
component then scales to an atomic gas column of $\rm
N_{H,\,SW\,atomic}\sim 2.4\e{21}\,cm^{-2}$.  Taking this SW peak as
indicative of the rest of the overall region, we conclude that the atomic
gas makes a small contribution to the total mass: $\sim$8\% of that of
the warm molecular material.
 
A plausible scenario is that the molecular and atomic gas are
physically associated and coexist as different phases in the same
morphological structures (e.g. clumps in the ring), as has been
proposed by \citet{gen85} and J93.  For clumps illuminated from one side
(as for a centrally-heated region centered on \sas), the fraction of photodissociated
material is estimated by the ratio of the extinction in a typical PDR layer
($\rm A_{V,\,PDR}\sim 4$) to the extinction through the full clump.
The fiducial 0.2~pc clumps / streamers discussed above would have $\rm
A_{V}=47$; thus for such structures, the mass ratio of the
photodissociated surface layer to the bulk is consistent with the mass
ratio based on the intensities.

\section{Heating of the Molecular Gas}
\label{sec:heating}

Having estimated the conditions and total cooling of the molecular gas
in the Galaxy's central 2~pc, we turn now to its heating source. There
are two potential mechanisms: photoelectric heating by UV photons, and
dynamical heating.  Both are a priori plausible near \sas, given its
intense UV sources and turbulent dynamical structure. Morphological
evidence suggests that UV heating may be important.  Warm CO peaks
interior to the HCN ring in general, and at the same position as the far-IR
emission in the southwest ring.  The emission observed toward the north of \sas\ may also be associated with the NA as
it enters the cavity.  Furthermore, the emission in
[SiII], [CII] and the far-IR continuum requires UV heating, and the
maps in these tracers are generally well-explained in the context of
$\sim 10^{7}$~\ls\ of visible-UV luminosity, most from near \sas,
illuminating dense clumps and streamers (L99).

However, a scenario that {\em only} involves UV photons exciting the molecular
gas is difficult to support quantitatively.  The
intensity of the CO \jseven\ transition relative to the atomic lines and
far-IR continuum is much larger than can be accounted for with any
existing PDR model.  The standard plane-parallel PDR models
\citep{th85,kau99} predict mid-$J$ CO emission arising from a layer
between $A_V\sim 3$ and $A_V\sim 7$: at $A_V\lesssim3$ CO is
photodissociated, and at $A_V\gtrsim7$ the UV does not produce the
heating necessary to excite these transitions.  These models predict a
[OI] 63\mm\ / CO \jseven\ ratio, $R$, between 600--1200 \citep{kau99},
nearly two orders of magnitude higher than our observed $R\sim15$.
The Orion Bar PDR presents a similar problem to the theory, though not
as extreme, with $R\sim130$ \citep{sta93}.  Clumpy models have been
generated in attempts to explain the discrepancy
\citep{bur90,mt93,kos94}, and \citeauthor{bur90} find that the Orion
PDR intensities can be reasonably well-modeled (to within factors of
two) as a mixture of clumpy PDRs, including some clumps with very high
densities ($\rm n= 10^7\ccm$).  Unlike in the Orion PDR, however, a
model which incorporates a large fraction of the gas at high density
is not supported for \sas, given the drop off in CO line intensity
above $J\sim 10$.  The best clumpy PDR model of \citeauthor{bur90} for
\sas\ based on the atomic lines, CO lines, and molecular hydrogen
lines underestimates the CO \jseven\ line intensity by a factor of 25.
The root of the problem is the fact that there is $\sim$15 times more
warm molecular gas than atomic gas, geometrically impossible for
UV-heated medium.  On this basis we rule out UV heating as a viable
method for all but a small fraction of the warm molecular gas we
observe.  \citet{rf01,rf04} reach a similar conclusion in their study
of \hh\ rotational line emission from other Galactic Center clouds.

\subsection{Shock Conditions in the Central 2~pc}
\label{shock}

We conclude that dissipation of supersonic turbulence in shocks is a
more likely source for heating the warm molecular material in the central 2~pc.
Dynamical heating has been proposed in explaining the emission from
rotational \hh\ lines in Galactic Center clouds \citet{rf04}, and the
presence of SiO in these clouds suggest a chemical history which
includes shocks \citep{hue98}.
Various models for the origin and dynamics of the features around \sas\ have been
constructed \citep[e.g.]{vd02,san98}, which include infalling clouds
and clump-clump collisions.  The presence of clumps which are not
Roche stable in the CND supports the notion that molecular material
feeds into the central parsecs, and that dynamical effects are
important.  Observations of molecular hydrogen provide further
evidence -- the molecular hydrogen ro-vibrational lines observed
toward the northeast and southwest CND show a spectrum which is almost
completely thermal, very much like the Orion shock
\citep{tan89,gat86}.  Qualitatively, the spectral tracers around \sas\
show some similarity to the shock region in Orion (BN-KL), with very
bright CO lines relative to all other tracers, [OI], [CII], and the
continuum (see Table~\ref{tab:lines}).  As indicated by the high-J
lines, however, the shock conditions in the bulk of the GC gas are clearly less
intense than the Orion KL shock, where the CO emission is modeled to
arise in the warm gas downstream of the 38\kms\ shock.
%

To derive the shock conditions, we invoke the magnetohydrodynamic
C-shock models of \citet[][hereafter DRD]{drd83}, \citet{dr84}, and
\citet{rd90}.  These models span a large parameter space---density,
velocity, and magnetic field all affect the post-shock conditions.
When applied to the GC region, the difficulty is the potentially
large magnetic field.  DRD present CO intensities only for
models with standard magnetic field, ${\rm B}=100\,\mathrm{\mu G}
\left( \frac{n} {10^4\ccm} \right)^{0.5}$.  The field strength has a
tremendous impact of the post-shock excitation, in general the peak
temperature is proportional to the square of the magnetic field and
square of the shock velocity.  An increase in the field strength by a
factor of 3 can reduce high-$J$ line intensities as much as a factor
of 100, while the peak $J$ and its intensity remain approximately
unchanged.  Other high-exitation lines such as those of \water, \hh,
and [OI] are similiarly affected.  The direct observational constraints on the field strength are
not strong because the Zeeman measurements are difficult.  It is known
that within 3--4 pc of \sas\ there are regions with B as large as 3 mG, though this does not pervade the region.  In
teh CND itself, for example, there is no measurable field with an
upper limit of 0.5 mG \citep{plc95,mly95}.  As these investigators
point out, however, that the large turbulent and orbital motions may distort and reverse the
field such that the average measured field in a large (44\as) beam may
underestimate its local strength.

Using the published low-magnetic-field models, the general shape of the CO line intensity
distribution, the peaking at $J$\s 8, and the falling-off toward higher $J$ is reproduced
for a range of density and velocity, from $n_{\rm H} \sim 10^4\ccm$, $v=20\,\kmsm$ to $n_{\rm
H} \sim 10^5\ccm$, $v = 13\,\kmsm$.  In all of these models cases, however, the transitions
below the peak (\s $J$= 1--5) are predicted to be brigher relative to the mid and high-$J$
transitions than is observed.  This is likely due in part to the lower-$J$ lines
being optically thin in the our observed region due to the large velocity dispersion, relative to the
models in which all of the lower-$J$ CO lines are optically thick.  A similar effect has been noted in considering the
ground-state (\jone) HCN transition emission observed the CND.  The large velocity dispersion produces a low optical
depth and reduces self-absorption, producing a high intensity per gas mass as compared with the more typical molecular
clouds with less motion (see e.g. \citet{wri01}).
Given this optical depth consideration, the higher density, lower velocity solution best reproduces the relative CO
intensities, underpredicting the $J\eqq8\!/\!J\eqq3$ intensity ratio by a factor of only 2. 
Large magnetic fields are appealing because they allow a great deal of energy to be radiated in a
fairly low-excitation gas, this is the sort of situation that is required to reproduce CO
luminosity that is comparable to or greater than the luminosity in [OI].  Given the CO
lines, the observed and inferred [OI] intensity,
and the published models of DRD, we conclude that a model with density of
10$^{4}$--10$^{5}\ccm$, and velocity-magnetic field product of 10--20 \kms\ \by\ 0.5--0.3 mG is the most likely
description of the shock conditions in the central 2~pc.  


In this scenario, the shock conditions which are
exciting the mid-$J$ CO are not likely the same as those which excite
the near-IR \hh\ lines.  \citet{gat86} have mapped a two-lobe
structure in the \hh\ $\nu\eqq1\too0\,S(1)$ around \sas, which is
consistent with the shape of the CND, and as pointed out above this
emission has been found to be due to shock excitation \citep{tan89}.
However, the low-velocity C-shocks which account for the observed CO
emission would not include appreciable luminosity in the near-IR \hh\
lines.  These scenarios are not inconsistent; there are likely a
variety of shock conditions around \sas.  \citet{gat86} claim that the
\hh\ emission is due to shocks at the inner edge of the molecular disk
due to mass outflow from \sas.  The CO emitting shocks, on the other
hand, arise from the dissipation of dynamical turbulence in the bulk
of molecular ring.

\subsection{Turbulent Dissipation and Orbit Decay in the Central 2~pc}
\label{sec:turb}

We now turn to the energetics of the turbulent dissipation.
\citet[and references therein]{sog98,ml99} have studied the
dissipation rate in magneto-hydrodynamic (MHD) turbulence, and find
that the rate is comparable to that of a pure hydrodynamical
turbulence, and is essentially given by dimensional analysis.  To get
the correct coefficient, we rearrange Equation~7 in \citet{ml99} to
express the energy output per mass of turbulent material as:

\begin{equation} 
\frac{L}{M} = 0.42 \frac{v_{\rm rms}^3}{\Lambda_d} = 1.10 \,
\left(\frac{v_{\rm rms}}{25\,  km\,s^{-1}}\right)^3 \left(\frac{1\,
\rm pc}{\Lambda_d}\right)\,\rm \frac{L_\odot}{M_\odot} , \label{eq:turb}
\end{equation}
where $v_{\rm rms}$ is the typical turbulent velocity, and $\Lambda_d$
is the size scale of the typical turbulent structures.  The turbulent
size scale should be comparable to or smaller than the smallest
observed clump size, of order 0.1~pc according to the interferometric
HCN measurements (also comparable to the Jeans length scale).  We
therefore adopt $\rm \Lambda_d\sim0.1\,pc$.  Taking a mid-range value
from the above analysis, we apply $\rm v_{rms} = 15\,\kmsm$ to
Eq.~(\ref{eq:turb}), yielding a power dissipated per unit mass of $\rm
(L/M)_{turb}=2.4$ in solar units.  This is a reasonable match to the
observed and inferred cooling rates in CO, [OI], and \hh.  Note also
that the adopted values for $\rm \Lambda_d$ and $\rm v_{rms}$ are
broadly consistent with the 300~\kmspc\ velocity gradient used in the
LVG analysis: $\rm 15\,\kmsm / 0.1\, pc \sim 150 km\,s^{-1}\,pc^{-1}$.

If the luminosity in the molecular lines is the result of turbulence
in the steady state, then the orbit of the molecular material must be decaying.
Differentiating the expression for Keplerian motion gives:

\begin{equation} 
\frac{dR}{dt} = \frac{2R^2}{GM} \cdot \frac{dE}{dt}, 
\end{equation}
where $M$ is the enclosed mass and $E$ represents the energy of per
mass of the molecular gas.  For an orbit around $\rm{4\e6
\,M_{\odot}}$ at a distance of 1.5 pc, the observed luminosity to mass
ratio corresponds to an orbit decay rate of $4\e{-6} \rm \,pc\,
yr^{-1}$.  Thus infall from the radius of the HCN ring (R$\sim$2~pc)
to the inner edge of the CND as traced in the far-IR (R$\sim$1.5~pc) would require $1.2\e5\rm
yr$, similar to the orbital timescale $\sim9\e4\,\rm
yr$.  Thus this gas is dissipating its orbital energy on a
orbit timescale, and cannot exist for more than a few orbits, or a few
$\rm \times10^5\, yr$.  This is not surprising if the material which
forms the CND and other features derives from orbits with low angular momentum, and the
CND itself represents the circularization of that material at a radius
appropriate for its angular momentum.
 
Recent modeling has given some theoretical basis for the formation of
the CND and streamers as the result of infalling clouds.
\citet{san98} produce an asymmetric elliptical structure, called a
dispersion ring, by tidal disruption of a clumpy cloud incident on a
point mass from 6.5 pc with angular momentum of a 1 pc radius circular
orbit.  After the material makes one or two passes (about $8\e5\,\rm
yr$), the resulting debris is a circumnuclear structure which
reproduces the kinematics of the HCN ring more accurately than any
circular rotation scenario, and hints at the same morphological
structure.
Though the overall distribution is not symmetric, the innermost edge of the dispersion ring is
elliptical, and consistent with the observed far-IR morphology
\citep{lat99}.  Throughout the simulation, the interior radius of the
CND gradually decreases, the result of repeated collisions between
clumps, which would certainly produce shocks and possibly trigger star
formation.  Whether the inner radius stabilizes, or some of the gas
evenutally accretes onto the central mass depends on the details of
the viscosity and the formation of stars.  \citet{san98} use a very
similar infalling cloud scenario to model the northern arm and the
east-west bar, each indendently.  This is appealing, as there is
convincing observational evidence that the northern arm is an
infalling streamer produced in a collision between two clouds
somewhere outside the CND \citep{lat99,her89,jac93}.  The northern arm
is in an earlier stage of infall than the CND gas, roughly on its
first passage, and has lower specific angular momentum, both of which
which make it more recognizable observationally.  It is significant
that the same model can generate the CND morphology and kinematics as traced with the far-IR and the HCN, and provide
a natural fit with the dynamically-driven energetics we derive for the material around \sas.

\section{Conclusions}
\label{sec:conclusions}

We present an 11\as\ spatial resolution map of the central 2 pc of the
galaxy in the \jj\ rotational transition of CO.  An LVG analysis
invoking all the available CO line intensities provides
constraints on the conditions in the molecular gas.  We find that
warm, 200--300 K, moderate density, $n \sim $5--7$\e4\ccm$, molecular gas
is abundant in the Galactic Center, with total CO luminosity
comparable to that of the atomic lines.  The total mass of warm
molecular material is at least 2000~\ms, slightly less than the total
mass measured with cooler molecular gas tracers, but an
order of magnitude more than the mass in photodissociated atomic gas.
Morphologically and kinematically, the CO emission bears some resemblance to the CND as traced with HCN and the
far-IR continuum.  Heating with a central UV source is not capable of
producing the observed CO luminosity, and we conclude that the primary energy source is dynamical.  The observed spectral
tracers are consistent with mechanical energy being dissipated into the gas in
low-velocity (10--20\kms) shocks with magnetic fields of 0.3--0.5 mG, and these values are consistent with other
observational evidence.
Our results support a scenario in which the CND and other features around \sas\ are each in the process of forming from
infalling molecular clouds with low angular momentum on timescales of 10$^{5}$ years.

\acknowledgments 
We thank Wayne Holland (currently at the UK ATC), Richard Prestage
(currently at NRAO Green Bank), and the JCMT operators and technical
staff for their help in making the commissioning run of SPIFI a
success.  We thank David Flower (Durham University) for providing the
high-temperature CO-\hh\ rate coefficients in advance of publication.
C.M.B. acknowledges support from a Milliken Fellowship at Caltech.  We thank an anonymous referee for several helpful
on a draft of the manuscript.

\clearpage

\begin{deluxetable}{cccccc}
\tabletypesize{\scriptsize}
\tablecaption{Galactic Center ISM tracers \label{tab:lines}}
\tablewidth{0pt}
\tablehead{}
\startdata
{} & \multicolumn{3}{c}{Galactic Center} & \multicolumn{2}{c}{Orion OMC 1}\\
{} & Sgr A$^*$\tablenotemark{a}&N. Arm (25\as)\tablenotemark{b}&SW CND (25\as)\tablenotemark{b}&Bar
PDR\tablenotemark{c}&KL Shock\tablenotemark{c}  \\
OI 63\mm &$2.1\e{-2}$&$4.1\e{-2}$&$2.3\e{-2}$&$4.0\e{-2}$&$2.7\e{-2}$\\ 
OI 145\mm &$8.\e{-4}$&\it{(1)}&\it{(0.93)}&$3.5\e{-3}$&$3.0\e{-3}$\\
CII 158\mm &$1.7\e{-3}$&\it{(1)}&\it{(1)}&$4.2\e{-3}$&$3.5\e{-3}$\\
CI 609\mm &$4.\e{-5}$& & & & \\
CO $J\eqq1\too0$& \multicolumn{3}{c}{------ $7.\e{-7}$ ------}&$4.\e{-7}$&$2.7\e{-6}$\\
CO $J\eqq7\too6$&$8\e{-4}$&$1.1\e{-3}$&$1.5\e{-3}$ & $3\e{-4}$ & $9\e{-3}$\\
CO $J\eqq14\too13$& $1.7\e{-4}$ & &  &$3.1\e{-4}$&$8.6\e{-3}$\\
Far-IR cont. &  22 & 50 & 26 & 26 & 390\\  Ratios & & & & &\\
\enddata
\tablenotetext{a}{Intensity ($\rm erg\,s^{-1}\,cm^{-2}\,sr^{-1}$) averaged over a 2\am\
 diameter beam around \sas.  The far-IR line intensities are taken from
 \citet{jac93,gen85}, the CO $J\eqq7\too6$ from this work, and the CO $J\eqq1\too0$ from
 \citet{ser86}.  The far-IR continuum is from the luminosity maps of \citet{lat99}, and
 includes a contribution of 30\% from the warm transiently heated dust grains which are
 not traced by the 32--38\mm\ continuum.  All intensities are corrected for extinction
 toward the galactic center, for which we take $\tau={23.5} \ {\lambda [{\mathrm \mu m}]}$
 after \citeauthor{lat99}.}  
\tablenotetext{b}{Average intensities over 25\as\ beams, for
 tracers with sufficient angular resolution, units and references as above.  Relative
 intensities in the two regions for tracers with larger ($\sim$1\am) beams are given in
 parentheses, estimated from the ISO LWS maps of \citet{whi98}.  (RA, Dec) Offsets
 from \sas: N. Arm, (+5\as, +15\as); SW CND, ($-$15\as, $-$30\as).}
\tablenotetext{c}{Intensities in \s 1\am regions centered on the far-IR bar, and the
 BN-KL object of Orion.  References are \citet{her97,wil86,sb89,sta93,wat85}.}
\end{deluxetable}

\clearpage


\begin{figure}
\caption[Integrated Intensity Map of CO $J\eqq7\too6$ in the Galactic Center
CND]{\label{fig:gc}Integrated intensity map of the CO $J\eqq7\too6$
emission from the Galactic Center Circumnuclear Disk (CND).  Offsets are relative to \sas\
($\rm{RA_{1950} = 17^h\,42^m\,29^s.3}$, $\rm{decl_{1950} =
-28^\circ\,59^{\prime}\,19^{\prime}}$), and the beam size is 11\as\ FWHM.  Emission is
observed throughout the the map; contours are linear in $\rm{T_{MB}\Delta v}$, with a
250\kkms\ interval.  The peak in the southwest CND is 5400\kkms, the minimum at the edges
of the map is 750\kkms.  Velocity resolution in the spectra is 70\kms.  Overall rotational
motion is evident in the shifting of the velocity from north to south.}
\end{figure}

\begin{figure}
\caption[CO $J\eqq7\too6$ Overlaid with Far-IR Continuum in the
CND]{\label{fig:overlayir}CO $J\eqq7\too6$ contours overlay the
greyscale far-IR continuum image of \citet{lat99}, which shows the morphological features
around \sas.  The southwest peak in CO is coincident with the far-IR peak in the southwest
CND, while the CO ridge to the north lies between the Northern Arm and the northwest CND.
The extension of the Northern Arm to the north is clearly detected in CO.  CO contours are
in units of $T_{\rm MB}\Delta v$, peaking at 5400\kkms, with interval 375\kkms.}
\end{figure}

\begin{figure}
\caption[CO $J\eqq7\too6$ Overlaid with HCN $J\eqq1\too0$ in the
CND]{CO $J\eqq7\too6$ contours overlay the greyscale
interferometric HCN \jone\ map \citep{gw00}.  CO generally peaks interior to the HCN, but shows the same
clumpy structure in the northern CND.  The gap in the HCN ring to the north corresponds to
the extension of the northern arm as traced in CO.  CO contours are as in
Figure~\ref{fig:overlayir}.\label{fig:overlayhcn}}
\end{figure}


\begin{figure}
\caption{LVG constraints for the gas in the CND.  Two line ratios are plotted:  \jseven\ to \jtwo\ (solid lines) and
\jseven\ to \jsixteen\ (dashed lines).   Heavy lines correspond to observed values in the SW CND, the two
medium-weight lines for each ratio show the range allowed by the 40\% uncertainty.   Calculations
assume a velocity gradient of 300\kmspc\ (see text), and the temperature range reflects the availability of CO-\hh\
collisional rate coefficients. \label{fig:gclvg}}
\end{figure}

\begin{figure} 
\caption[LVG Fit to CO Intensities in the CND]{Model of the excitation and radiative transfer of CO using a large-velocity-gradient, escape-probability approximation.  Observed intensities are averaged over
a 44\as\ beam centered at ($\delta$ RA = -15\as, $\delta$ Dec = -30\as) from \sas.  CO
$J\eqq7\too6$ is from this work, the other points from \citet{lug87,ser86,sut90}.  Error
bars are conservative at 30\%, reasonable for the calibration uncertainties in the
submillimeter and far-IR lines, but likely an overestimate of the uncertainty in the
lower-J lines.  The lighter curves demonstrate intensities that would be generated in LTE
and without including the effects of radiative excitations of the upper levels.  \label{fig:cofit}}
\end{figure}

\clearpage

\includegraphics{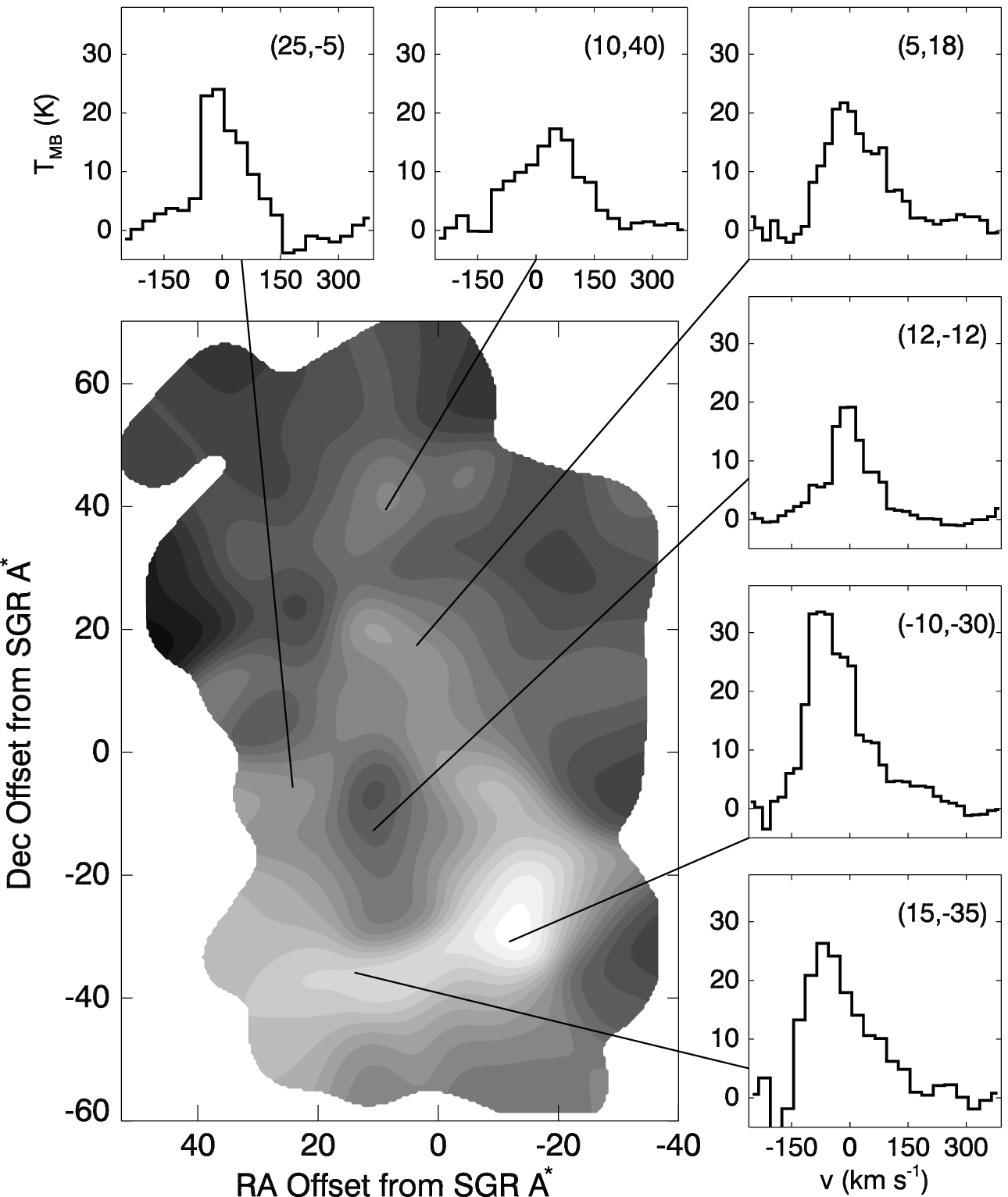}

Figure 1, Bradford et al.

\clearpage

\plotone{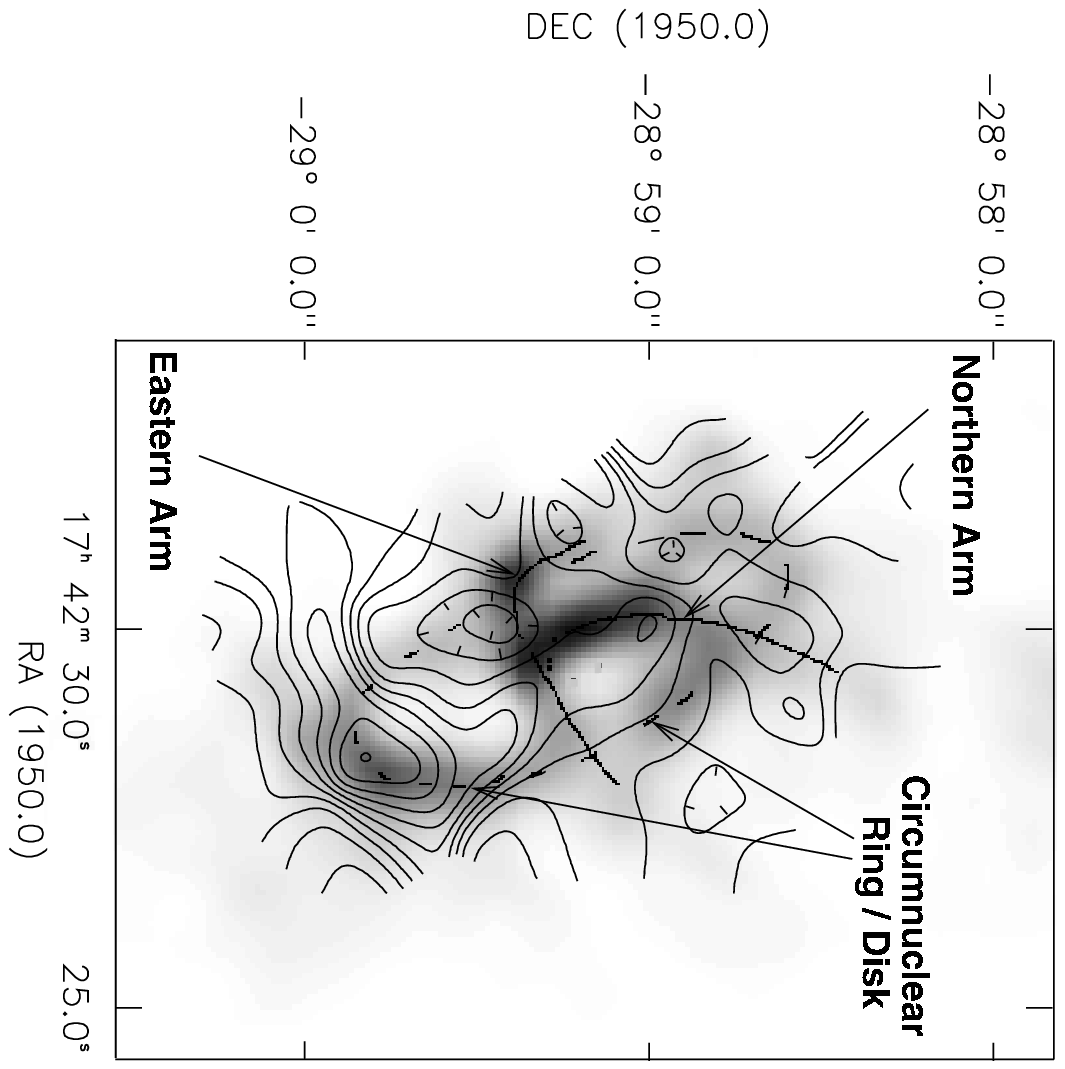}
Figure 2, Bradford et al.

\clearpage

{\rotatebox{0}{\includegraphics{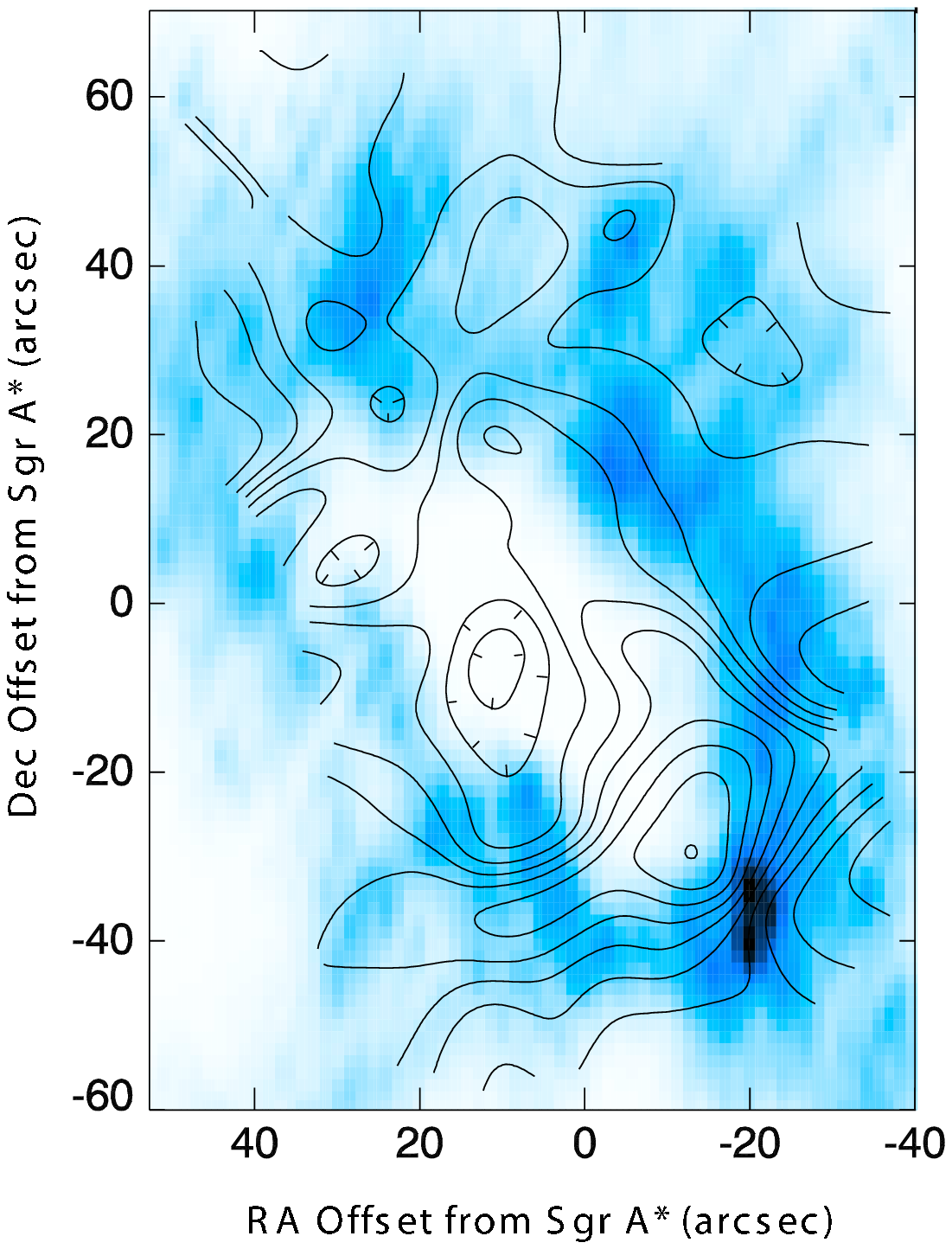}}}\\

Figure 3, Bradford et al.

\clearpage \plotone{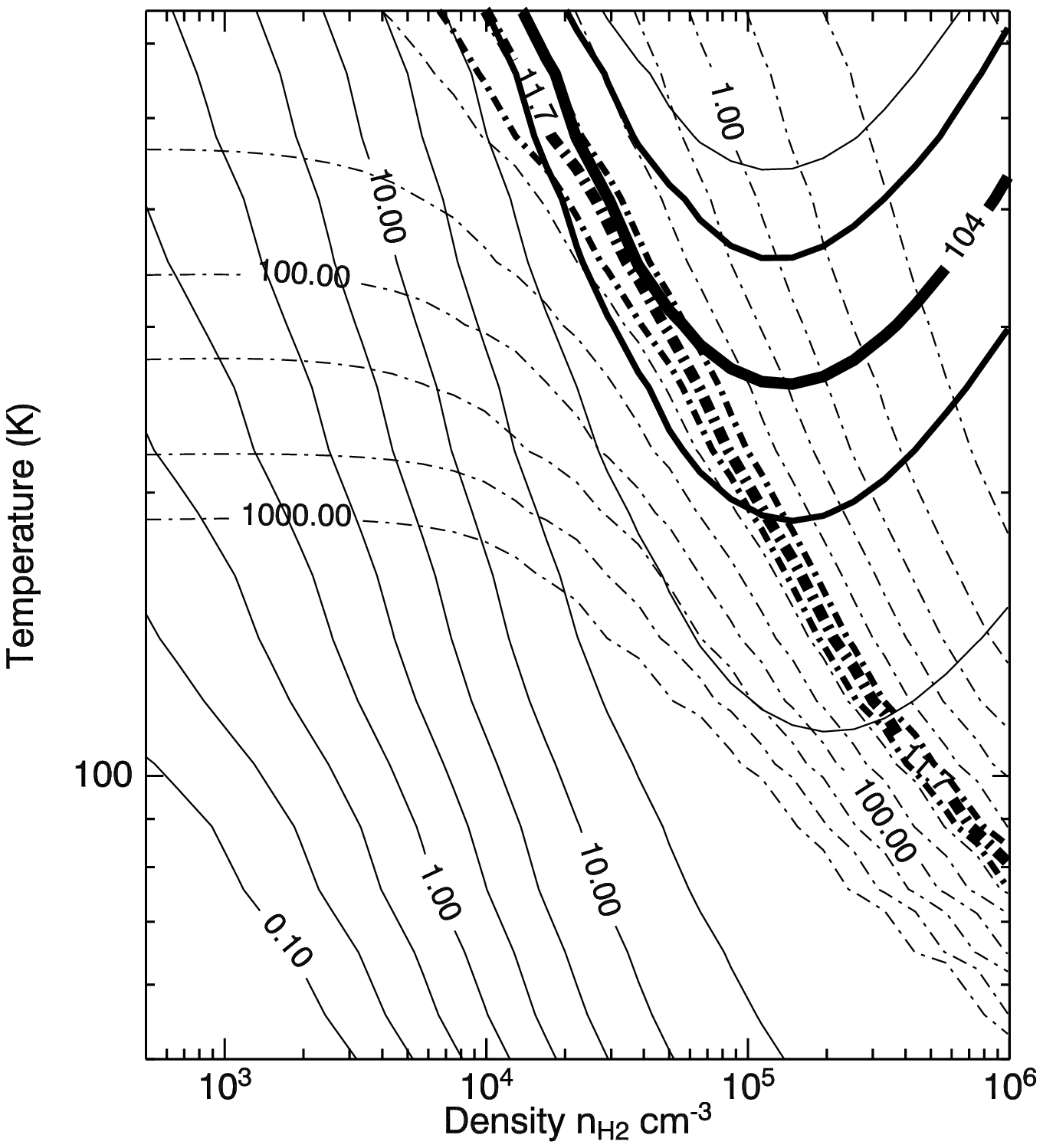}

Figure 4, Bradford et al.

\clearpage \plotone{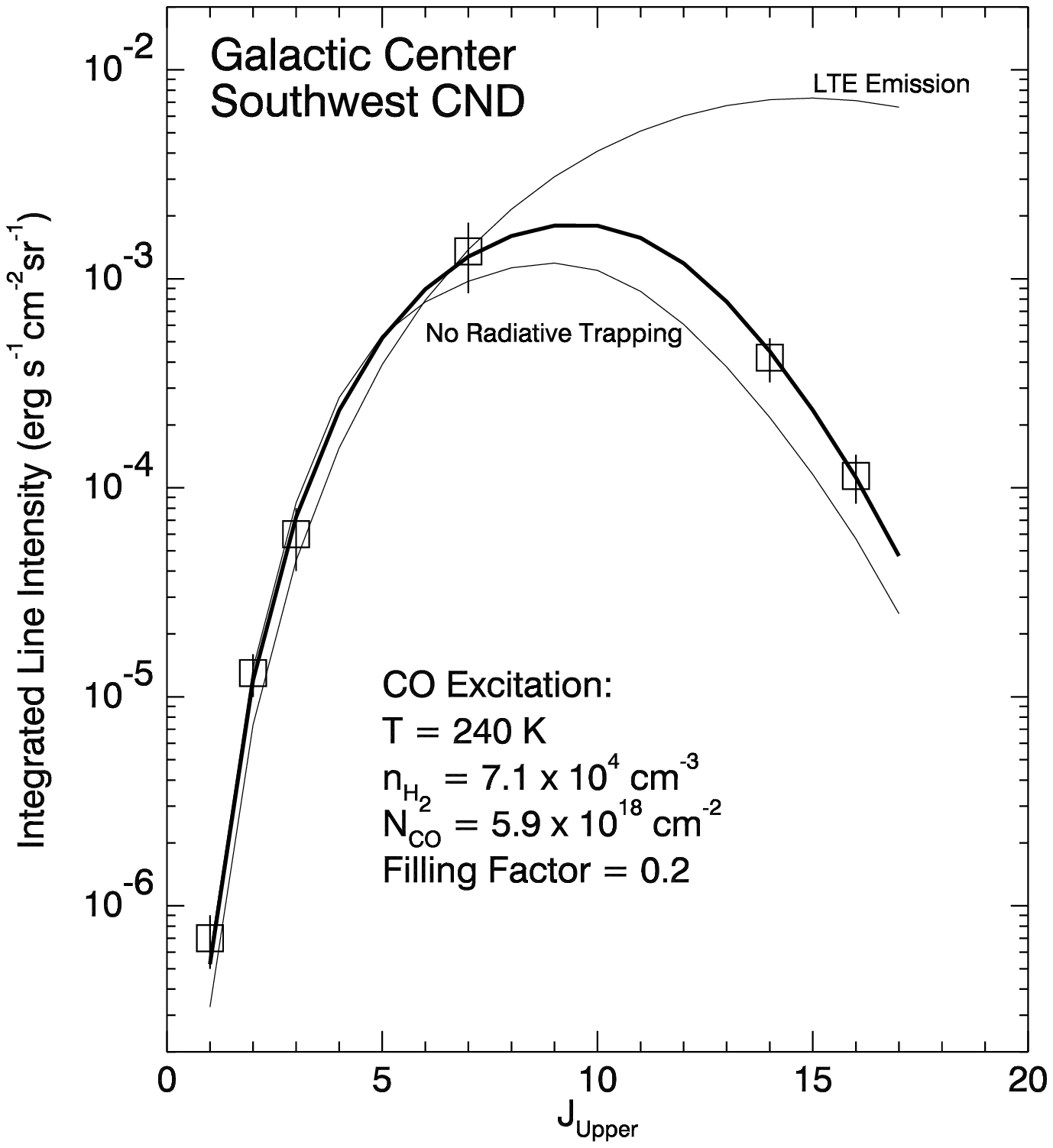}

Figure 5, Bradford et al.

\end{document}